\documentclass[aps,prl,superscriptaddress,reprint]{revtex4-2}
\usepackage{graphicx}
\usepackage{hyperref}
\usepackage{amsmath}
\usepackage{color}
\usepackage{newtxtext,newtxmath}
\definecolor{myblue}{rgb}{0,0,1}

\begin{document}

\title{\large Reinforcement learning-guided optimization of critical current in high-temperature superconductors}

\author{Mouyang Cheng}
\email{vipandyc@mit.edu} 
\altaffiliation{Equal contribution}
\affiliation{Quantum Measurement Group, MIT, Cambridge, MA 02139, USA}
\affiliation{Center for Computational Science and Engineering, MIT, Cambridge, MA 02139, USA}
\affiliation{Department of Materials Science and Engineering, MIT, Cambridge, MA 02139, USA}

\author{Qiwei Wan}
\altaffiliation{Equal contribution}
\affiliation{Quantum Measurement Group, MIT, Cambridge, MA 02139, USA}
\affiliation{Department of Nuclear Science and Engineering, MIT, Cambridge, MA 02139, USA}

\author{Bowen Yu}
\altaffiliation{Equal contribution}
\affiliation{Quantum Measurement Group, MIT, Cambridge, MA 02139, USA}
\affiliation{Department of Physics, MIT, Cambridge, MA 02139, USA}

\author{Eunbi Rha}
\affiliation{Quantum Measurement Group, MIT, Cambridge, MA 02139, USA}
\affiliation{Department of Nuclear Science and Engineering, MIT, Cambridge, MA 02139, USA}

\author{Michael J Landry}
\affiliation{Quantum Measurement Group, MIT, Cambridge, MA 02139, USA}
\affiliation{Department of Nuclear Science and Engineering, MIT, Cambridge, MA 02139, USA}
\affiliation{Department of Physics, MIT, Cambridge, MA 02139, USA}

\author{Mingda Li}
\email{mingda@mit.edu} 
\affiliation{Quantum Measurement Group, MIT, Cambridge, MA 02139, USA}
\affiliation{Center for Computational Science and Engineering, MIT, Cambridge, MA 02139, USA}
\affiliation{Department of Nuclear Science and Engineering, MIT, Cambridge, MA 02139, USA}

\begin{abstract}
High-temperature superconductors are essential for next-generation energy and quantum technologies, yet their performance is often limited by the critical current density ($J_c$), which is strongly influenced by microstructural defects. Optimizing $J_c$ through defect engineering is challenging due to the complex interplay of defect type, density, and spatial correlation. Here we present an integrated workflow that combines reinforcement learning (RL) with time-dependent Ginzburg–Landau (TDGL) simulations to autonomously identify optimal defect configurations that maximize $J_c$. In our framework, TDGL simulations generate current–voltage characteristics to evaluate $J_c$, which serves as the reward signal that guides the RL agent to iteratively refine defect configurations. We find that the agent discovers optimal defect densities and correlations in two-dimensional thin-film geometries, enhancing vortex pinning and $J_c$ relative to the pristine thin-film, approaching 60\% of theoretical depairing limit with up to 15-fold enhancement compared to random initialization. This RL-driven approach provides a scalable strategy for defect engineering, with broad implications for advancing HTS applications in fusion magnets, particle accelerators, and other high-field technologies.

\end{abstract}

\maketitle

Superconductivity has long been regarded as a ``holy grail'' of condensed matter physics, with high-temperature superconductors (HTS) holding particular promise for large-scale applications \cite{zhou2021high,coombs2024high}. 
While much of the field has historically focused on increasing the superconducting transition temperature ($T_c$), practical considerations limit the utility of this pursuit. For example, many record-high $T_c$ materials require extreme pressures exceeding 100 GPa, precluding their mass deployment in real-world environments \cite{somayazulu2019evidence,drozdov2015conventional,drozdov2019superconductivity}. 
In contrast, cooling with liquid nitrogen at 77 K, adequate for many HTS, remains relatively inexpensive and practical in laboratory and industrial settings. Furthermore, with modern power transmission technologies already exceeding 95\% efficiency in the U.S. \cite{hausman2025power}, the near-term energy advantage of further increasing $T_c$ might be less pronounced. Consequently, optimizing properties beyond $T_c$ is increasingly important for scalable superconducting technologies.

Beyond $T_c$, another decisive factor for scalable applications of HTS is the critical current density ($J_c$). This quantity defines the maximum current density a material can carry while remaining superconducting, thereby setting both the minimum feasible device dimensions and the upper limits of operable magnetic fields. 
A high $J_c$ is essential not only for compact superconducting electronics but also for large-scale systems such as fusion reactors \cite{ibekwe2025use,cohen2024long,coombs2024high}, particle accelerators \cite{padamsee2001science,rossi2012superconducting}, and high-field magnets for medical imaging \cite{parizh2016conductors,manso2023superconducting}. Although progress in superconductivity has been largely driven by the quest to raise $T_c$ from a fundamental perspective, advancing $J_c$ remains equally crucial for translating these materials into practical with significant efforts \cite{larbalestier2001high,foltyn2007materials,gutierrez2007strong,gurevich2011use,glatz2020quest,ruiz2025critical,puig2024impact}. Unlike $T_c$, which depends on microscopic pairing mechanisms and requires detailed electronic-structure understanding, $J_c$ can be modeled within the mean-field Ginzburg–Landau framework, making it more reliably predictable and experimentally accessible.
It is well established that $J_c$ can be substantially tuned through defect engineering, particularly by tailoring vortex-pinning landscapes. Within the time-dependent Ginzburg–Landau (TDGL) framework, the controlled introduction of defects, such as nanoinclusions and dislocations, can enhance pinning efficiency and increase $J_c$ \cite{schmid1966time,larkin1979pinning,dew1974flux,kramer1978theory,wang2017parallel,blair2018time}.
Despite substantial experimental progress in optimizing $J_c$ in HTS \cite{selvamanickam2015high,coll2014size,majkic2018over,maiorov2009synergetic,macmanus2004strongly}, the search for optimal defect landscapes remains inefficient and could be greatly improved or accelerated through computationally guided approaches.
From a computational perspective, recent advances in parallelized TDGL algorithms have enabled large-scale simulations \cite{sadovskyy2015stable,jonsson2022current,bishop2023pytdgl}, supporting optimization strategies based on evolutionary and particle-swarm algorithms \cite{kimmel2017silico,sadovskyy2019targeted}. Nevertheless, the vast combinatorial space of possible defect configurations remains a major challenge for systematic $J_c$ optimization.

In this Letter, we introduce an integrated framework that combines deep reinforcement learning (RL) with TDGL simulations to optimize defect configurations in an HTS thin film. 
RL, in which an agent interacts with an environment and refines its strategy by maximizing expected cumulative rewards, has achieved landmark successes such as self-learning the Atari games \cite{mnih2013playing} and mastering Go through AlphaGo \cite{silver2017mastering}. 
Drawing on this analogy, we cast defect engineering as a high-dimensional optimization game, where the RL agent ``plays'' by proposing defect configurations as actions and receives the $J_c$, computed from TDGL current–voltage ($I\text{-}V$) characteristics, as the reward. 
Through this iterative process, the agent autonomously identifies optimal defect configurations including defect densities, spatial correlations, and arrangements that enhance vortex pinning for a given sample geometry and superconductor characteristics. The optimized configuration yields an increase of $J_c$ approaching $\sim0.6J_\mathrm{dp}$, corresponding to up to 15-fold enhancement relative to a randomly initialized defect configuration. While this improvement appears moderate compared with some experimental enhancements, it should be interpreted in the context of the clean, two-dimensional TDGL model used here, which excludes many extrinsic factors that can further raise $J_c$ in experiments. Within such an idealized mean-field environment, reaching 60\% of the depairing current represents a near-saturation regime where additional gain is physically constrained, indicating that the RL agent has approached the intrinsic optimization limit under the simulated conditions. Our RL-guided framework thus establishes a generalizable platform for autonomous defect optimization of $J_c$, with broad implications for scalable HTS technologies and other quantum systems where defect control is critical.

\begin{figure}[!htbp]
  \centering
  \includegraphics[width=0.5\textwidth]{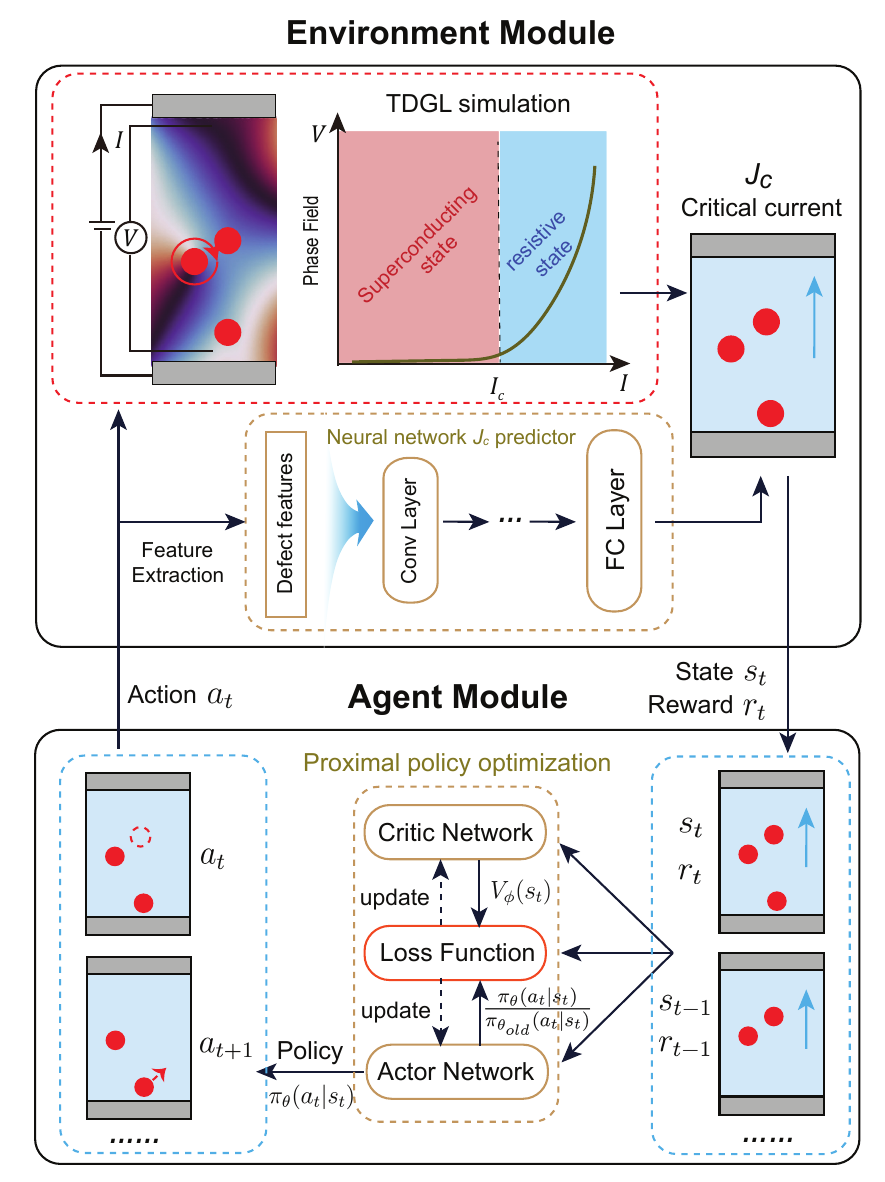}
  \caption{Overview of deep reinforcement learning (RL) framework for optimizing the critical current $J_c$. The Environment Module (top panel) evaluates the $J_c$ for defect configurations proposed by the agent using time-dependent Ginzburg-Landau (TDGL) equation. A surrogate machine learning model is concurrently trained to directly predict the $J_c$, reducing the need for repeated TDGL simulations. The Agent Module (bottom panel) performs defect engineering by selecting actions to update the defect configuration via proximal policy optimization (PPO) with an actor-critic network.}
  \label{fig1}
\end{figure}

The workflow of the RL-based optimization of $J_c$ is shown in Fig.\,\ref{fig1}. It consists of two modules: an Environment Module that evaluates $J_c$ for each defect configuration proposed by the agent using the TDGL formalism, and an Agent Module, which updates defect configurations to increase $J_c$.
The simulated system consists of a two-dimensional HTS thin film with open boundary conditions. Two opposite sides are connected to electrodes that drive a current flow $I$, and the voltage $V$ across the sample is computed using the TDGL equation, offering a mean-field description of the evolution of the superconducting order parameter $\psi=\psi(\mathbf{r},t)$ \cite{schmid1966time,kramer1978theory}.
In dimensionless units, the TDGL equation can be written as \cite{bishop2023pytdgl}
\begin{equation}
\begin{aligned}
\frac{u}{\sqrt{(1+\gamma^2|\psi|^2)}}\bigg(\frac{\partial}{\partial t}+i\mu &+\frac{\gamma^2}{2} \frac{\partial |\psi|^2}{\partial t}\bigg)\psi \\ &=(\epsilon-|\psi|^2)\psi+(\nabla-i\mathbf{A})^2\psi
\end{aligned}
\label{TDGL}
\end{equation}
where the spatial-temporal dependencies $(\mathbf{r},t)$ are omitted for brevity. Here $\mu$ is the electric scalar potential, $\mathbf{A}$ is the vector potential, $\epsilon(\mathbf{r})=T_c(\mathbf{r})/T-1$ characterizes the local critical temperature. The parameter $u=\pi^4/14\xi(3)\approx5.79$, and $\gamma=1$ parameterize the relaxation dynamics and the inelastic scattering, respectively. When coupled with the Poisson equation
\begin{equation}
\nabla^2\mu=\nabla\cdot\text{Im}[\psi^*(\nabla-i\mathbf{A})\psi]-\nabla\cdot\frac{\partial\mathbf{A}}{\partial t},
\label{Poisson}
\end{equation}
the TDGL formalism can self-consistently determine the evolution of both the electric scalar potential $\mu$ and the order parameter $\psi$ in space and time.

We construct a two-dimensional sample of size $32\xi \times 16\xi$ and thickness $d=0.1\xi$, where $\xi$ denotes the superconducting coherence length.
Defects are modeled as metallic nanoinclusions with various diameters $2\xi, 4\xi$ and $6\xi$, which are implemented by setting $\epsilon(\mathbf{r})=-1$ within the defect regions to represent normal metal state, and $\epsilon(\mathbf{r})=+1$ elsewhere to represent the superconducting state. 
The resulting $I\text{-}V$ characteristics for arbitrary defect configurations are computed using the pyTDGL package \cite{bishop2023pytdgl}. 
The critical current density $J_c$ is then obtained by monitoring the point at which the voltage exceeds a prescribed threshold. More details of the TDGL simulation can be found in the \textit{End Matter}.

In this RL framework, the state $s_t$ represents the current defect configuration of the thin film.
By interacting with the environment via a reward $r_t$, namely the TDGL-evaluated $J_c$ at iteration $t$, the agent progressively refines its policy $\pi_\theta(a_t|s_t)$ for proposing new defect configurations, where the action $a_t$ involves displacing, adding, or removing a defect. We employ the proximal policy optimization (PPO) algorithm, a widely used actor--critic method that stabilizes training by clipping the policy update ratio $w=\pi_\theta/\pi_{\theta_{\text{old}}}$ to prevent overly large gradient updates. The actor network generates actions $a_t$, corresponding to an update of the defect configuration, while the critic network estimates the value function $V_\phi(s_t)$ to assess the expected return from state $s_t$. Both networks are jointly updated through the PPO clipped surrogate loss function:
\begin{equation}
L(\theta) = \mathbb{E}_t \Big[ 
\min \big( w_t(\theta) \hat{A}_t,\; 
\text{clip}(w_t(\theta), 1-\epsilon, 1+\epsilon) \hat{A}_t \big) \Big],
\end{equation}
where $w_t(\theta) = \pi_\theta(a_t|s_t)/\pi_{\theta_{\text{old}}}(a_t|s_t)$ is the policy ratio, $\epsilon$ is the clipping parameter, and $\hat{A}_t$ is the estimated advantage evaluated using the generalized
advantage estimation (GAE) method \cite{schulman2015high}. 
Importantly, the clipping mechanism ensures that defect updates remain incremental, discouraging abrupt configuration changes that could destabilize the TDGL solver. Further implementation details of PPO are provided in the \textit{End Matter}.

Through repeated interaction with the TDGL environment, the agent autonomously identifies defect densities, spatial correlations, and pinning arrangements that maximize vortex pinning efficiency and thereby enhance $J_c$ up to 0.6 $J_\text{dp}$, where $J_\text{dp}$ is the theoretical maximum current density predicted by the TDGL theory. This RL-driven framework moves beyond heuristic, trial-and-error defect engineering, offering a scalable route to optimize superconducting $J_c$ under complex, high-dimensional design space.

Direct calculation of $J_c$ at each RL update $s_t$ is computationally prohibitive: a typical training run of $10^5$ iterations would require $10^5$ full TDGL simulations. 
Such simulations are expensive because they rely on implicit Euler integration to obtain stable and converged vortex dynamics, which becomes particularly time-consuming near the critical regime close to $J_c$. 
To overcome this bottleneck, we introduce a surrogate model that evaluates $J_c$ directly from the defect configuration (top panel of Fig.\,\ref{fig1}), replacing explicit TDGL evaluations at every iteration. This surrogate greatly accelerates the RL process, with minimal loss in accuracy. The final optimized $J_c$ is then validated by a smaller set of full TDGL evaluations, ensuring both efficiency and reliability.

\begin{figure}[!htbp]
  \centering
  \includegraphics[width=0.5\textwidth]{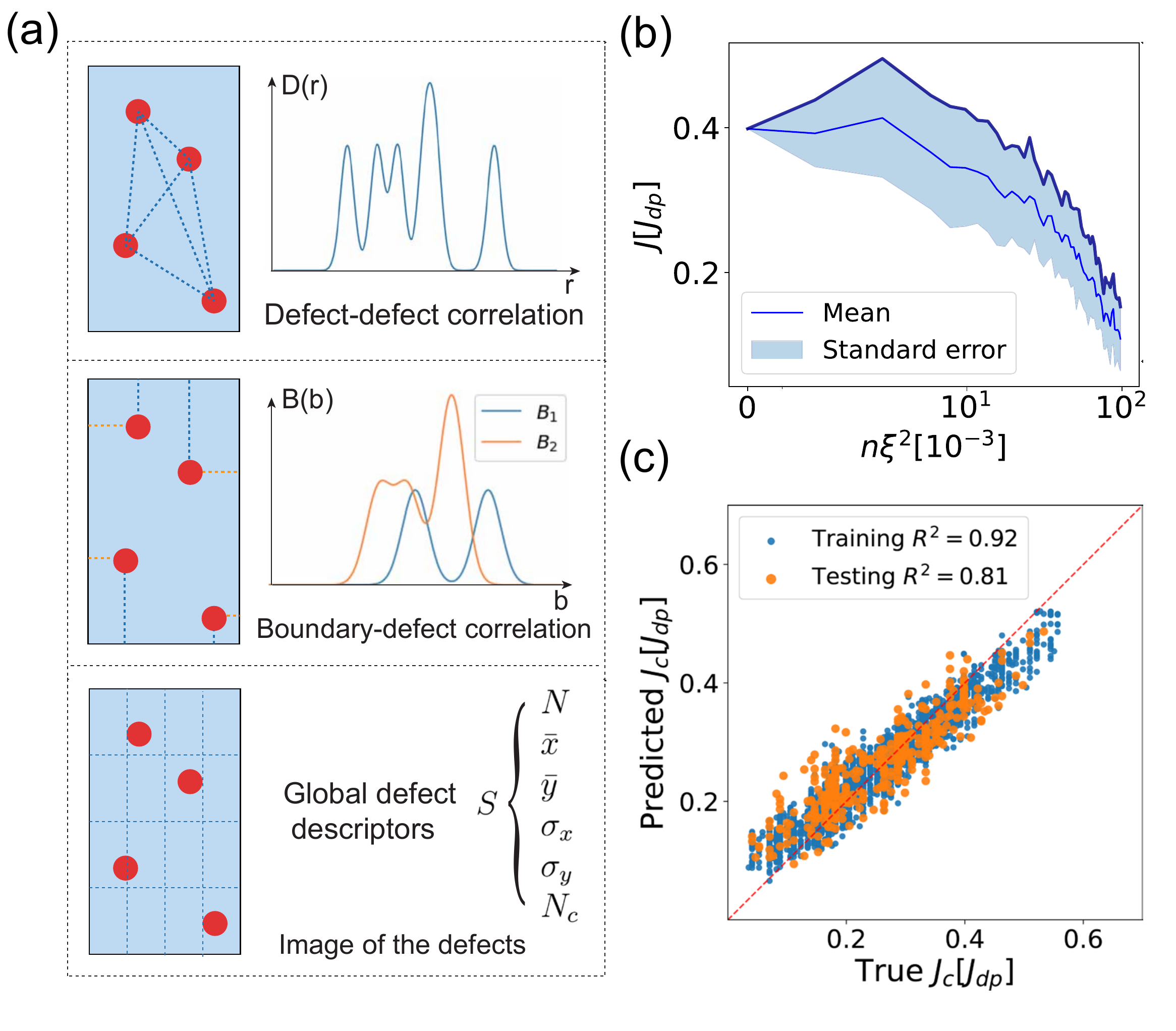}
  \caption{Machine learning prediction of $J_c$ in HTS thin films. \textbf{(a)} Physics-informed defect descriptors used for model training, including defect-defect pair correlation $D(r)$ (top), boundary-defect correlation $B(b)$ (middle), global descriptors for defect geometry and concentration (bottom). \textbf{(b)} Distribution of simulated $J_c$ with increasing defect concentration $n$ with same diameter $d=2\xi$. The shaded blue area denotes the range of the standard error. Here $\xi^2$ stands for coherence length of the HTS thin film in our simulation. \textbf{(c)} Correlation plot for machine learning prediction of $J_c$ with defect diameter $d=2\xi$, where both training and testing results are shown.}
  \label{fig2}
\end{figure}

The results of the surrogate ML model for predicting $J_c$ from defect configurations are shown in Fig. \ref{fig2}, using defect diameter $d=2\xi$ as a representative case.
To reduce the high-dimensional defect configuration input $s_t$ into physics-aware low-dimensional representation, we construct a set of defect-related descriptors for ML training (Fig. \ref{fig2}(a)). These descriptors are designed to capture the key mechanisms underlying vortex pinning and dynamics. For instance, the defect--defect pair correlation function $D(r)$ reflects the degree of clustering or uniformity among defects, which strongly influences the collective pinning landscape experienced by vortices. Likewise, the boundary--defect correlation $B(b)$ encodes the proximity of defects to film edges, a factor that can modify vortex entry and exit behavior. Global descriptors summarize coarse-grained information such as the total defect number $N$, average position $(\bar{x}, \bar{y})$, spatial variance $(\sigma_x, \sigma_y)$, and fraction of defects within the central region $N_c$, thereby providing compact statistical measures of defect geometry and density. 

For machine learning, we generate 3,000 HTS thin films with distinct defect configurations. Details on the parameter choices are provided in \textit{End Matter}.
Fig.\,\ref{fig2}(b) shows the dependence of the simulated $J_c$ on defect concentration $n$, spanning nearly two orders of magnitude in dimensionless parameter $n\xi^2$. Intuitively, at low concentrations, the introduction of defects enhances $J_c$ by providing additional pinning centers that suppress vortex motion. 
Beyond an optimal density, increasing disorder leads to vortex channeling, reduced coherence, and ultimately the suppression of superconductivity. This non-monotonic behavior is accompanied by substantial variation in $J_c$ at a fixed defect concentration $n$, indicating that defect clustering, alignment, or proximity to boundaries can further alter the local pinning landscape and vortex dynamics. 
These observations motivate our choices of defect descriptors and highlight the necessity of a data-driven ML predictor capable of efficiently predicting $J_c$.

The $J_c$ predictor is implemented as a six-layer convolutional neural network (CNN) \cite{lecun1998convolutional}. In this architecture, the physics-informed defect descriptors shown in Fig.\,\ref{fig2}(a) are first processed through separate CNN layers, then concatenated and passed through a multi-layer perceptron (MLP) backbone to produce a unified prediction of $J_c$. The model is trained using the Adam optimizer with learning rate $\eta=5\times 10^{-4}$.
Fig.\,\ref{fig2}(c) illustrates the performance of the surrogate ML model in predicting $J_c$. The model achieves high accuracy with training and testing $R^2$ scores of 0.92 and 0.81, respectively, indicating that the selected physics-informed descriptors effectively capture the key defect structures that govern the vortex pinning and $J_c$. 
Feature-importance analysis further reveals the boundary correlation $B$ as the leading importance in the $J_c$ prediction, at least for a reasonably small simulation box.

\begin{figure}[!htbp]
  \centering
  \includegraphics[width=0.35\textwidth]{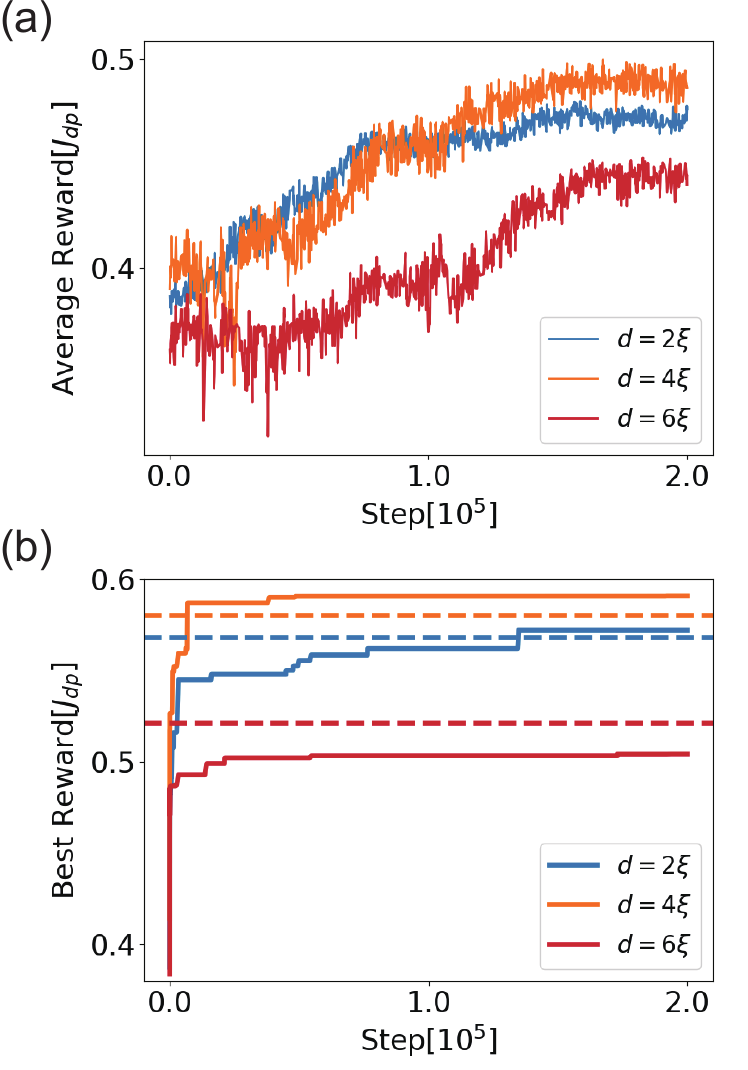}
  \caption{Reinforcement learning (RL) optimization of the critical current density $J_c$ for different defect diameters. \textbf{(a)} Evolution of the average of the most recent 100 rewards $J_c/J_{\mathrm{dp}}$ as a function of training steps for defect diameters $d = 2\xi$, $4\xi$, and $6\xi$. \textbf{(b)} Best-achieved $J_c$ as a function of training steps for defect diameters $d = 2\xi$, $4\xi$, and $6\xi$. The dashed lines mark the final validation using full TDGL simulations, whereas the solid curves represent the RL optimization guided by the surrogate model.}
  \label{fig3}
\end{figure}

Having established how defect features influence $J_c$ and constructed an efficient surrogate $J_c$ predictor, we next train the RL model for on-the-fly optimization of $J_c$. Such on-the-fly capability can be directly applied to \textit{in-situ} guide defect introduction experiments, such as ion implantation, or to real-time monitoring of fusion reactor operation for autonomous decision-making. 
To examine the defect size dependence of $J_c$, on defects, we consider three types of metallic nanoinclusions with diameters $d=2\xi, 4\xi$ and $6\xi$. For each defect size, a separate $J_c$ predictor is trained on 3,000 defect configurations labeled by TDGL, followed by $2\times10^5$ steps of RL optimization with PPO algorithm. Each RL run begins from a randomly initialized defect configuration on the HTS thin film.

The results are summarized in Fig.\,\ref{fig3}. As shown in Fig.\,\ref{fig3}(a), the average reward, defined as $J_c/J_{\mathrm{dp}}$, steadily increases with training, indicating progressive improvement of the defect configurations discovered by the agent. 
The best-achieved rewards in Fig.\,\ref{fig3}(b) show that the optimized $J_c$ is enhanced by approximately 40\% relative to the pristine sample. 
The dashed lines denote the true TDGL-evaluated $J_c$, which closely follow the surrogate predictions at convergence. This validates the robustness and generalizability of the surrogate model, even when the state space $s_t$, representing the defect configuration, extends beyond the training data distribution.
Furthermore, $J_c$ exhibits a strong dependence on the defect diameters: the intermediate diameter $d = 4\xi$ yields the largest enhancement, approaching $J_c/J_{\mathrm{dp}} \approx 0.6$. This result indicates an optimal balance between enhanced vortex pinning and minimal suppression of the superconducting order parameter.
Moreover, compared to the randomly initialized defect configuration which could have $J_c$ as low as $\sim 0.04 J_\text{dp}$, our RL workflow is capable of enhancing $J_c$ by up to 15 times.

To summarize, we have developed a workflow that integrates deep RL with TDGL simulations to optimize $J_c$ of HTS thin films via defect engineering. 
Our results show substantial (up to 15-fold) increase of $J_c$, approaching 60\% of the theoretical depairing limit. Despite the success, several challenges remain.
First, the scope of our simulation is limited to circular-shaped, mono-disperse nanoinclusions. Real HTS may host diverse microstructural defects, including dislocations, grain boundaries, and voids, whose interplay can further complicate the vortex pinning with further $J_c$ enhancement. 
Second, this study employs a two-dimensional thin-film geometry. While this choice improves numerical stability and computational efficiency, it inevitably neglects three-dimensional effects such as vortex line bending and defect correlations along the $z$-axis. Extending the current framework to full three-dimensional TDGL simulations would better capture these effects and provide a richer action space for exploring defect-driven optimization of $J_c$.
Additionally, thermal fluctuation is neglected in this study, and numerical setup such as the finite size of the simulation box, the choice of applied magnetic field and $\epsilon(\mathbf{r})$ may cause discrepancies from simulations to experimental observations. These factors may limit the direct quantitative comparison with experimental data, though the qualitative trends and optimization insights remain robust.
Finally, while the RL framework successfully identifies optimal defect configurations computationally, experimental realization remains challenging. Recent development on ion accelerator may offer a possible pathway for such defect engineering \cite{mandal2023accelerator}.
Establishing a mapping between the agent’s design proposals and practical experimental controls, such as ion implantation parameters, will be crucial to closing the loop between simulation and experiment.

Even with those caveats, our results highlight the power of RL for guiding defect optimization. The RL policy captures the generalizable optimization strategies rather than fixed configurations, achieving efficient and stable improvement through the actor–critic framework and clipped policy updates using the PPO algorithm.
Combined with deep neural networks, this approach enables more efficient exploration of the high-dimensional defect state space, resulting in a higher optimal $J_c$ than heuristic baselines such as evolutionary methods or particle-swarm optimization \cite{kimmel2017silico,sadovskyy2019targeted}. Moreover, a surrogate $J_c$ model accelerates the environment–agent feedback cycle without compromising ranking fidelity, allowing broader exploration before performing high-fidelity TDGL evaluations. 
Combining the surrogate model with RL training, once our RL model is properly trained, it allows on-the-fly optimization of $J_c$ using only 256 surrogate $J_c$ evaluations and merely a few TDGL evaluations. This advanced framework is much more efficient than traditional algorithms, which typically requires more than 6,000 TDGL evaluations \cite{sadovskyy2019targeted} (see justification in \textit{End Matter}).
Taken together, our work establishes a scalable, physics-informed framework for inverse design and defect engineering. The same strategy can be readily extended to other quantum and semiconducting platforms, where controlling defects within high-dimensional design spaces is essential for optimizing performance and realizing unconventional phenomena, paving the way toward advanced defect-tunable quantum technologies.


\section{Acknowledgments}
The authors thank SE Ferry, ZS Hartwig, DG Whyte, P Hirschfeld, R Hennig, B Geisler, and J Hamlin for the helpful discussions. MC and MJL acknowledges support from U.S. Department of Energy (DOE), Office of Science (SC), Basic Energy Sciences Award No. DE-SC0021940 and DE-SC0020148. BY and ER thank support from National Science Foundation (NSF) ITE-2345084. ML acknowledges the support from MIT Energy Initiative Future Energy System Center (FUSC) program and support from R. Wachnik. 

\bibliography{refs.bib} 

\section{End Matter}
\textit{Time-dependent Ginzburg-Landau simulation.}
The time-dependent Ginzburg-Landau (TDGL) equation provides a mean-field description of the evolution of the superconducting order parameter $\psi(\mathbf{r},t)$, allowing the study of the voltage-current characteristic and the vortex motion.\cite{schmid1966time,kramer1978theory} Since the equations Eq. \ref{TDGL} and \ref{Poisson} are written in dimensionless form, physical quantities such as the coherence length and the London penetration length do not appear explicitly.
In this work, we construct a two-dimensional sample of size $32\xi \times 16\xi$ with the thickness $d=0.1\xi$, the London penetration length $\lambda=4\xi$ and the magnetic field $B=0.1B_{c2}$ applied perpendicular to the sample, where $\xi$ is the coherence length and $B_{c2}=\frac{h}{4\pi e \xi^2}$ is the critical magnetic field. 

To obtain the voltage-current characteristic, the electrodes are attached to the ends of the sample to inject a current, and the voltage response is computed by solving the TDGL equation via the Python package pyTDGL. The differential equations are discretized using a finite-volume scheme with a maximum grid size of $\xi/2$ to ensure accuracy, and the discretized equations are then solved using the implicit Euler method \cite{bishop2023pytdgl}. The critical current density $J_c$ is then determined by setting the threshold voltage $V_\text{th}=0.02V_0$, where $V_0=\frac{4\xi^2B_{c2}}{\mu_0\sigma\lambda^2}$ is the voltage unit in pyTDGL and $\sigma$ is the normal state conductivity, and applying a bisection algorithm. As the voltage approaches this threshold, the onset of vortex depinning can be observed, which validates the selection of the threshold voltage. Throughout the paper, the critical current density $J_c$ is expressed in units of the depairing current density $J_\text{dp}=\frac{h}{6\sqrt{3}\pi e\mu_0\lambda^2\xi}$.

\textit{Details about $J_c$ prediction.}
In generating the dataset for model training, the number of defects 
$N$ is randomly chosen between 0 and 50, and the defects are randomly placed within the sample. The model input consists of the features extracted from the configuration of the defects, including $D(r)$, $B_{1,2}(b)$, $S$ and $G$ shown in Fig. \ref{fig2}. Here, $D(r)$ represents the fraction of defect pairs separated by a distance $r$; $B_1(b)$ and $B_2(b)$ denote the fraction of defects whose nearest distances to the horizontal and vertical boundaries are $b$ respectively; $S$ includes global descriptors such as the total number of defects $N$, average coordinates $\bar{x},\bar{y}$, coordinate variances $\sigma_x,\sigma_y$ and fraction of defects in the central region $N_c$, and $G$ is the image of the defect distribution. All these features are discretized and normalized before being used as the input, and the output, the critical current density $J_c$, is also normalized. Features $D, B_{1,2}, G$ are individually processed through several layers of convolutional layers to extract spatial representations, wheras the scalar feature $S$ is transformed by fully-connected layers. The outputs from all branches are subsequently concatenated and fed into three fully connected layers to yield the final prediction. The architecture of the neural networks in the reinforcement learning algorithm is identical to that in the prediction model. 

\begin{figure}[!htbp]
  \centering
  \includegraphics[width=0.48\textwidth]{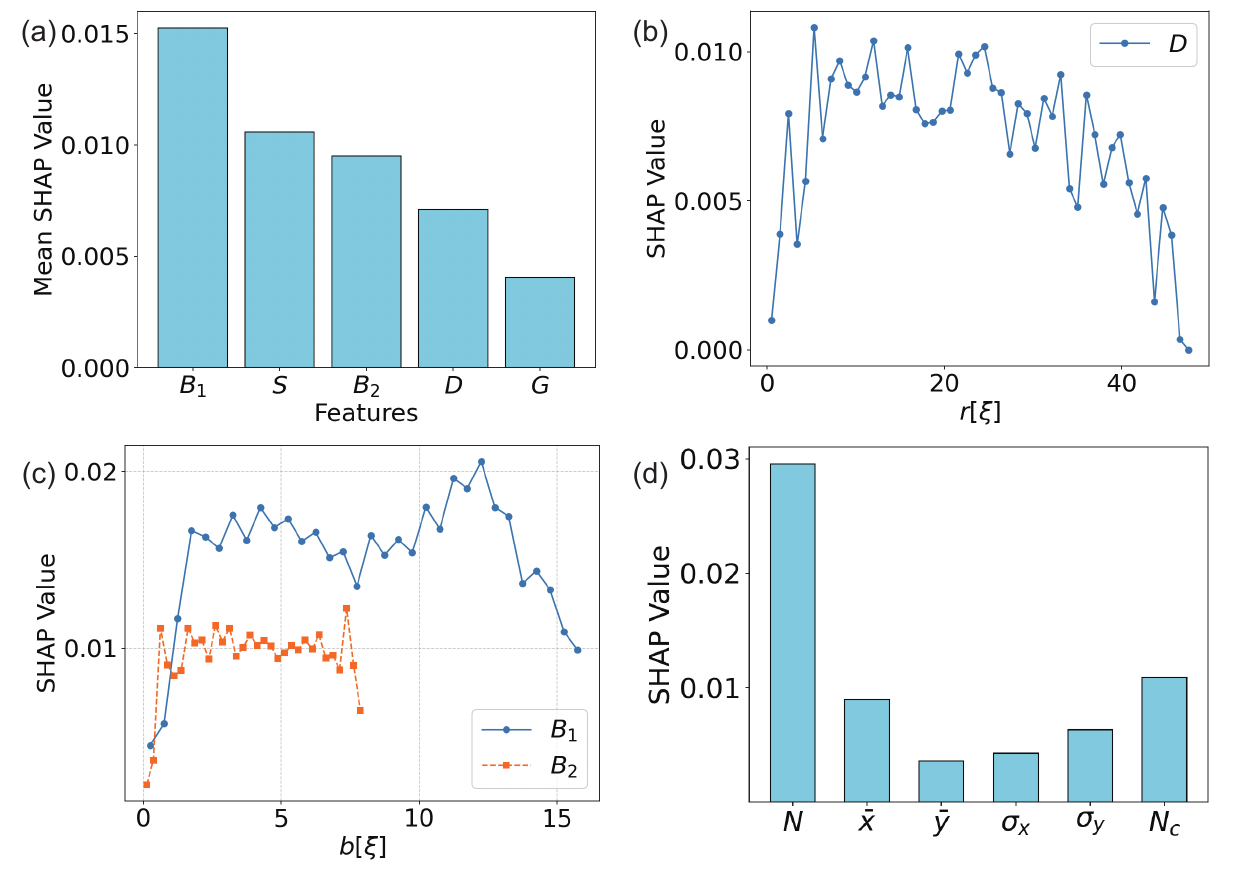}
  \caption{Shapley Additive Explanation (SHAP) analysis of feature importance in the RL optimization model. \textbf{(a)} Mean SHAP values for correlation descriptors ($B_1(b)$, $B_2(b)$, $D(r)$), global descriptors ($S = \{N, \bar{x}, \bar{y}, \sigma_x, \sigma_y, N_c\}$), and the entire image ($G$), averaged over all training samples. \textbf{(b)} SHAP value distribution for the correlation descriptor $D(r)$. \textbf{(c)} SHAP value distributions for boundary–defect correlation descriptors $B_1(b)$ and $B_2(b)$. \textbf{(d)} SHAP value components corresponding to global descriptors.}
  \label{fig4}
\end{figure}

To show the contribution of each input feature to the model prediction, we employed the Shapley Additive Explanation (SHAP) framework \cite{lundberg2017unified}. SHAP values provide a measure of feature importance based on the cooperative game theory, attributing to each feature its marginal contribution to the model output. The results shown in Fig. \ref{fig4} illustrate that $B_1(b)$ features contribute the most to the model’s output, followed by $S$, $B_2(b)$ and $D(r)$; the least efficient feature is the entire image $G$. As lower and more compact representations of $G$, features $B_1(b)$, $S$, $B_2(b)$ and $D(r)$ have demonstrated their abilities to encode the most important features of the image more efficiently than the big block of image. Furthermore, the results shown in Fig. \ref{fig4}(c), in line with the summary provided in Fig. \ref{fig4}(a), suggest that the defect distribution along the vertical direction may play a more important role. Finally, Fig. \ref{fig4}(d) evaluates a different set of summary statistics features ($N$, $\bar{x}$, $\bar{y}$, $\sigma_x$, $\sigma_y$, and $N_c$), with $N$ showing dominant influence and $N_c$ also standing out. This suggests that compared to the detailed distribution of defects, the total number of defects is more imperative in determining the critical current.

\textit{Deep reinforcement learning setup.}
We use proximal policy optimization (PPO) to iteratively maximize the critical current $J_c$ \cite{schulman2017proximal}. PPO is an actor–critic method in which an agent (\textit{actor}) learns by trial and error under the guidance of a learned \textit{value} estimator (\textit{critic}) \cite{konda1999actor}. In our setup, the superconducting system serves as the environment that receives an action $a_t$ (removing, adding, or displacing defects) and returns the next state $s_t$ and reward $r_t = R(s_t, a_t)$, namely the resulting critical current $J_c$.

The policy $\pi_\theta(a|s)$ defines the agent’s probabilistic action-selection strategy. In PPO, it is modeled as a diagonal Gaussian distribution over the continuous action space:
\begin{equation}
\pi_\theta(a | s)=\mathcal{N}\!\big(a;\,\mu_\theta(s),\,\mathrm{diag}(\sigma_\theta^2)\big),
\end{equation}
where the neural network outputs the mean $\mu_\theta(s)$ and the logarithm of the standard deviation $\log\sigma_\theta$. 

The cumulative reward under a specific policy $\pi(a|s)$ is described through the $Q$-function, $Q^\pi(s,a)=\mathbb{E}_\pi\!\left[\sum_{t=0}^{T-1}\gamma^{t}R(s_t,a_t)\mid s_0{=}s,a_0{=}a\right]$, estimating the expected discounted return starting from state $s$ and action $a$, and the value function, $V^\pi(s)=\mathbb{E}_{a \sim \pi(\cdot|s)}[Q^\pi(s,a)]$ averaging over all possible actions. In our setup, the discount factor is set as $\gamma=0.99$, and $V(s)$ is parameterized as $V_\phi(s)$ by processing each input component through separate convolutional neural networks (CNNs) or linear branches, concatenating the outputs, and feeding them into an MLP.

During each training iteration $k$, the agent resets the environment to initial conditions, collects rewards $r_t^k$ through the trained surrogate model, and generates a trajectory $\tau_k=\{(s_t^k,a_t^k,r_t^k)\}_{t=0}^{T-1}$ with $T=256$ roll-out steps under the current policy $\pi_{\theta_k}$. To evaluate the loss, we utilize the advantage function $A_t^k\equiv Q^\pi(s_t^k,a_t^k)-V^\pi(s_t^k)$, which measures how much better the action $a_t^k$ is compared with the expected average action at the state $s_t^k$. This can be estimated by generalized advantage estimation (GAE) \cite{schulman2015high}:
\begin{equation}
    \hat{A}_t^k = \sum_{l=0}^{T-t-1} (\gamma \lambda)^l \delta_{t+l}^k, \quad \delta_t^k = r_t^k + \gamma V_{\phi_k}(s_{t+1}^k) - V_{\phi_k}(s_t^k)
\end{equation}
with $\lambda=0.95$. The empirical return is thus $R_t^k=V_{\phi_k}(s_t)+\hat{A}_t^k$.

At the end of the iteration, both the policy and value networks are jointly optimized. The policy parameters $\theta$ are updated to maximize the expected advantage along the collected trajectories, while the value parameters $\phi$ are updated to minimize the prediction error between the estimated $V_\phi(s_t^k)$ and the empirical returns $R_t^k$ calculated from GAE. The update is weighted by the importance ratio $w_t^k(\theta)=\pi_\theta(a_t^k|s_t^k)/\pi_{\theta_k}(a_t^k|s_t^k)$, clipped within $[1-\epsilon,1+\epsilon]$ with $\epsilon=0.2$ to prevent overly large updates. The joint objective is
\begin{equation}
\theta_{k+1},\phi_{k+1}=\arg\min_{\theta,\phi}\Big[L_{\mathrm{p}}^k(\theta)+c_{\mathrm{v}}L_{\mathrm{v}}^k(\phi)\Big],
\end{equation}
where
\begin{equation}
\begin{aligned}
    L_{\mathrm{p}}^k(\theta)&=-\mathbb{E}_{\tau_k}\!\left[\min\!\big(w_t^k\hat{A}_t^k,\;\mathrm{clip}(w_t^k,1-\epsilon,1+\epsilon)\hat{A}_t^k\big)\right], \\
    L_{\mathrm{v}}^k(\phi)&=\mathbb{E}_{\tau_k}\!\big[(V_\phi(s_t^k)-R_t^k)^2\big],
\end{aligned}
\end{equation}
and we set $c_{\mathrm{v}}=0.5$ to balance the two loss terms. The total number of interactions with the defect environment is $N = 2\times10^5$, corresponding to about $N/T \approx 781$ training iterations. Under this setup, once training is complete, the learned policy can efficiently identify the optimal defect configuration within 256 steps of rapid evolution and requires only a single final TDGL evaluation. In contrast, traditional approaches, such as evolutionary algorithms or particle-swarm optimization, typically demand over 6,000 TDGL evaluations, which would be much more time-consuming than our RL-based model.

\end{document}